\documentclass[twocolumn,superscriptaddress,nofootinbib]{revtex4-2}

\usepackage{amsmath,amssymb}
\usepackage{graphicx}
\graphicspath{{figures/}{./}}
\usepackage{booktabs}
\usepackage[dvipsnames]{xcolor}
\usepackage{hyperref}

\newcommand{\Mpl}{M_\mathrm{pl}}
\newcommand{\wde}{w_\mathrm{de}}
\newcommand{\LCDM}{\Lambda\mathrm{CDM}}
\newcommand{\class}{\textsc{class}}
\newcommand{\cobaya}{\textsc{cobaya}}
\newcommand{\phiini}{\phi_\mathrm{ini}}
\newcommand{\szwelve}{\sigma_{12}}

\begin{document}

\title{Can coupled thawing quintessence reproduce DESI DR2?\\
       A test with two potentials}

\author{Jhananii Yuvaraj Babu}
\affiliation{Institut f\"ur Theoretische Physik, Universit\"at Heidelberg,
Philosophenweg 16, D-69120 Heidelberg, Germany}

\author{Luca Amendola}
\affiliation{Institut f\"ur Theoretische Physik, Universit\"at Heidelberg,
Philosophenweg 16, D-69120 Heidelberg, Germany}
\affiliation{Center for Astrophysics and Space Science, NYUAD Research Institute, New York University Abu Dhabi, PO Box 129188, Abu Dhabi, UAE}

\date{\today}

\begin{abstract}

DESI DR2 favours dark energy whose equation of state evolves crossing
$w=-1$ in the past. Coupled quintessence can produce such an apparent phantom
crossing without a phantom field, since a coupled dark sector analysed under
the assumption of separate conservation yields an effective equation of state
below $-1$. We test whether this mechanism can reproduce the DESI result,
using two thawing potentials, a cosine and a double-exponential hilltop,
confronted with DESI DR2 BAO, Planck PR4 and three supernova compilations.
Both models improve on $\LCDM$ by $\Delta\chi^2 = 5$--$15$, recovering
$64$--$92\%$ of the improvement achieved by $w_0w_a$CDM. They track the
$w_0w_a$CDM equation of state at $z\lesssim0.4$ but cross the phantom divide
later, near $z\simeq1$ rather than $z\simeq0.4$--$0.5$.  The coupling is bounded at
$|\beta|<0.073$ ($95\%$), and the two potentials agree on it to within
$0.15\sigma$. They differ instead in the equation of state they can reach: the
cosine comes closer to the evolution preferred by DESI and is weakly preferred
by the fit. The coupling posterior is bimodal, with modes at
$\beta\simeq\pm0.05$; only the negative mode produces the crossing, and it is
the higher of the two in every analysis. In conclusion,  the data favour the coupled quintessence phantom
crossing but do not yet require it.
\end{abstract}

\maketitle

\section{Introduction}
\label{sec:intro}

The second data release of the Dark Energy Spectroscopic Instrument
\cite{DESIDR2}, combined with cosmic microwave background and supernova
data, prefers a dark-energy (DE) equation of state that evolves with redshift.
Fitted with the Chevallier-Polarski-Linder (CPL) parametrisation
$w(a) = w_0 + w_a(1-a)$, the data exclude $\LCDM$ at roughly the $3\sigma$
level, and the preferred parameters imply $w(z) < -1$ in the past. Similar
indications have emerged from model-independent reconstructions of the
expansion history \cite{Calderon2024,Lodha2025}.

A genuinely phantom dark-energy field is theoretically hard to explain: a
canonical scalar with $w<-1$ requires a mathematically reversed sign in its
kinetic energy term \footnote{More in general, denoting with $K(\phi,X)$ the Lagrangian kinetic term, where $X\equiv - \phi^{;\mu}\phi_{;\mu}/2$ and $\phi$ is the scalar field,  what is needed for a phantom equation of state is $0<d\log |K|/d\log X<1/2$. }, creating a ``ghost field'' with negative kinetic energy.
It has long been shown, however, that an \emph{apparent} phantom phase can
arise without any such pathology if the dark sector is coupled
\cite{Sola2005,Gannouji2006,Das2006}. If dark matter (DM) and DE exchange
energy, but an observer analyses the data assuming the two are separately
conserved, the effective equation of state that observer infers can cross the
phantom divide even though the underlying field never does. Coupled
quintessence \cite{Wetterich1988,Wetterich1995,Amendola2000}, in which the dark matter mass
depends on a light scalar field, explains this mechanism with a single
additional parameter, the coupling strength $\beta$.

This possibility has attracted considerable attention since the DESI results
appeared, and a number of groups have confronted coupled dark energy with the
DR2 data \cite{Chakraborty2026,Li2026,deSouza2026,Wang2025,GomezValent2026}.
The reported constraints, however, disagree substantially. Significances for a
non-zero coupling range from roughly $2\sigma$ \cite{GomezValent2026} to more
than $5\sigma$ \cite{Li2026}, and the analyses do not agree on how to treat the sign of
$\beta$. Refs.~\cite{Chakraborty2026,Li2026,deSouza2026,Wang2025} all fix it
through a one-sided prior, $\beta\ge0$ in their own conventions (the coupling
of Ref.~\cite{Chakraborty2026} enters with the opposite sign in the exponent to
ours, so its prior corresponds to $\beta\le0$ here), and only
Ref.~\cite{GomezValent2026} explores both signs, in separate chains. A
preference for one sign cannot be inferred from a one-sided analysis. The
analyses also differ in the dark-energy sector: Ref.~\cite{GomezValent2026},
for instance, adopts a flat potential and a Peebles-Ratra power law, while the
others use different forms of the potential or of the interaction.

The question that motivates this work is: can coupled quintessence
reproduce the DESI result? Answering it with a single potential leaves open
whether any success or failure belongs to the coupling or to the particular
chosen potential.We therefore test two thawing potentials, in which the field
is frozen at $w\simeq-1$ by Hubble friction until late times and only then
begins to roll, and ask both whether each is able to reproduce the DESI
preference and how the result differs between them.

We address this by confronting two thawing potentials with identical data,
using the same code, the same priors on shared parameters and the same
treatment of nuisance parameters throughout. The two are chosen to be as
dissimilar as the thawing requirement allows: a cosine
(pseudo-Nambu-Goldstone) potential, which is periodic and characterised by an
amplitude and a decay constant; and a double-exponential hilltop
characterised by two logarithmic slopes,  which we adopt following a
suggestion of C.~Wetterich based on Ref.~\cite{Wetterich2024}. They differ in geometry, in parameter content,
and in the numerical treatment their closure conditions require
(Sec.~\ref{sec:method_shooting}). Agreement between them is
therefore a non-trivial test rather than a reparametrisation.

In brief, both potentials recover most of the improvement over $\LCDM$
achieved by the $w_0w_a$CDM fit, and the constraint on the coupling is the same
for both, while the equation of state they can reach is not. The coupling
posterior is bimodal, and only its negative mode produces the phantom crossing.

The paper is organised as follows. Section~\ref{sec:model} sets out the model,
the origin of the apparent phantom crossing, and the two potentials.
Section~\ref{sec:method} describes the data and the sampling.
Section~\ref{sec:results} gives the constraints and compares the two
potentials.
Section~\ref{sec:future} discusses what future measurements can and cannot
do, and Sec.~\ref{sec:literature} compares with previous work. We conclude in
Sec.~\ref{sec:conclusions}.

\section{Coupled thawing quintessence}
\label{sec:model}

In thawing quintessence the field is frozen by Hubble friction in the early
universe, where $H$ greatly exceeds its effective mass, so that its potential
energy acts as a cosmological constant and $w\simeq-1$. As $H$ falls the
friction weakens, the field starts to roll down its potential and acquires
kinetic energy, and $w$ rises from $-1$ toward less negative values. This is
the direction of the DESI preference for $w>-1$ today. For a canonical field
$w$ never falls below $-1$, so the apparent phantom crossing studied here
arises only through the coupling (Sec.~\ref{sec:model_crossing}).

In uncoupled thawing quintessence the potential matters surprisingly little:
as long as the field starts frozen near $w=-1$ and rolls only slowly, the
late-time equation of state follows a nearly universal trajectory fixed by
$w_0$ alone, largely independent of the shape of the potential $V(\phi)$~\cite{ScherrerSen2008,
DuttaScherrer2008}. With a DM-DE coupling this is no longer guaranteed. The
apparent phantom crossing derived in Sec.~\ref{sec:model_crossing} is set by
the field displacement $\Delta\phi$, and how far and how fast the field rolls
is a property of the potential. The potential also decides which starting
values are compatible with today's dark energy density, and therefore which
region of parameter space the data can explore at all. A single potential
therefore cannot tell us whether a successful (or failed) fit to DESI is a
property of the coupling or of the particular $V$ chosen. We address this by
analysing two potentials of very different shape, described next.

\subsection{Two thawing potentials}
\label{sec:model_potentials}

\begin{figure}[t]
  \centering
  \includegraphics[width=\columnwidth]{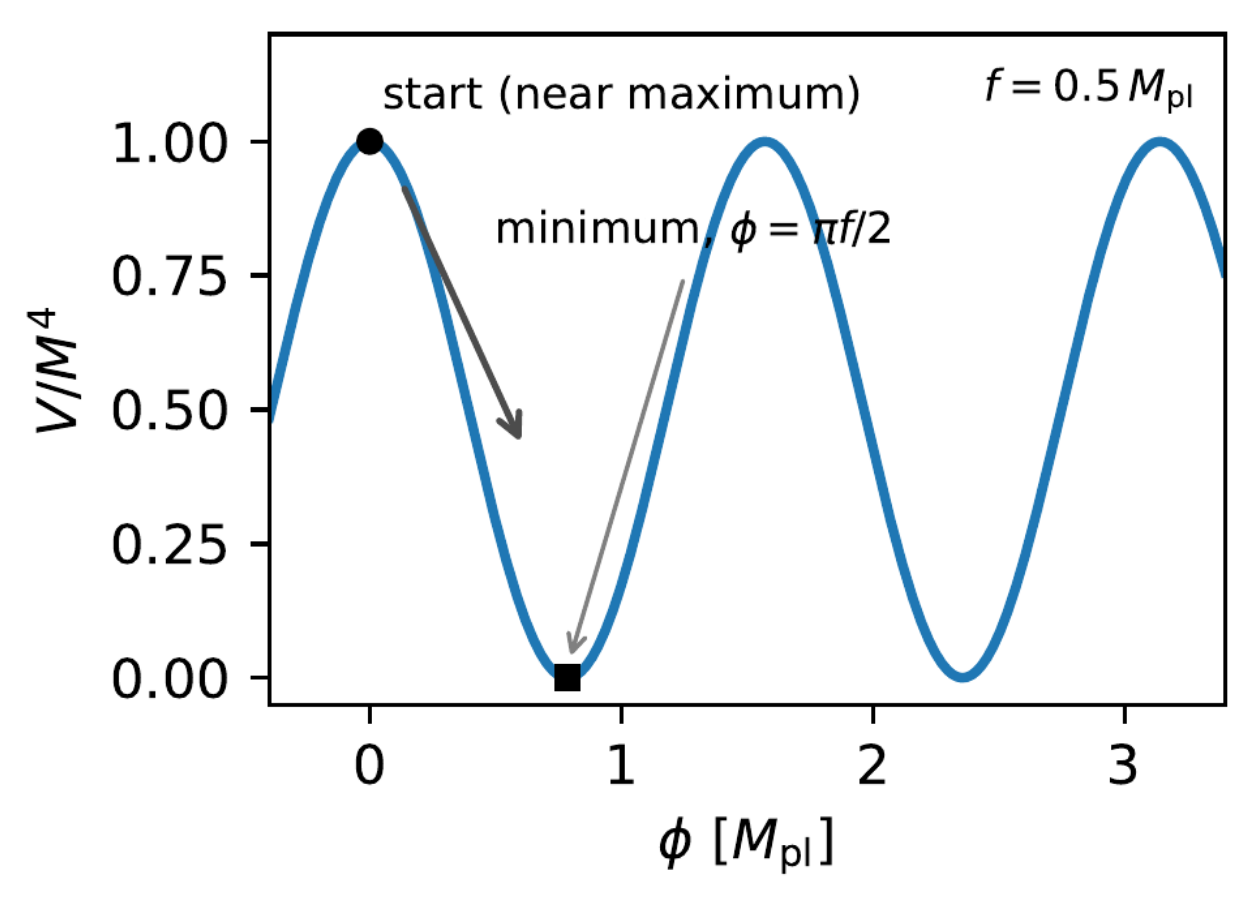}
  \caption{Shape of the cosine potential $V(\phi)=M^4\cos^2(\phi/f)$, shown
  for $f=0.5\,\Mpl$, with the field starting near the maximum and rolling
  toward the first minimum.}
  \label{fig:pot_cosine}
\end{figure}

The cosine (pseudo-Nambu-Goldstone) potential,
\begin{equation}
  V(\phi) = M^4\cos^2\!\left(\phi/f\right),
  \label{eq:cosine}
\end{equation}
is periodic, with maxima at $\phi = n\pi f$ where $V=M^4$, and minima at
$\phi = (n+\tfrac{1}{2})\pi f$ where $V=0$. Its shape is shown in
Fig.~\ref{fig:pot_cosine}. In the thawing scenario the field is initialised
near a maximum, where the potential is flat and the field is frozen; as $H$
falls the field is released and rolls down toward the neighbouring minimum,
and it is during this roll that $w_\phi$ departs from $-1$.

\begin{figure}[t]
  \centering
  \includegraphics[width=\columnwidth]{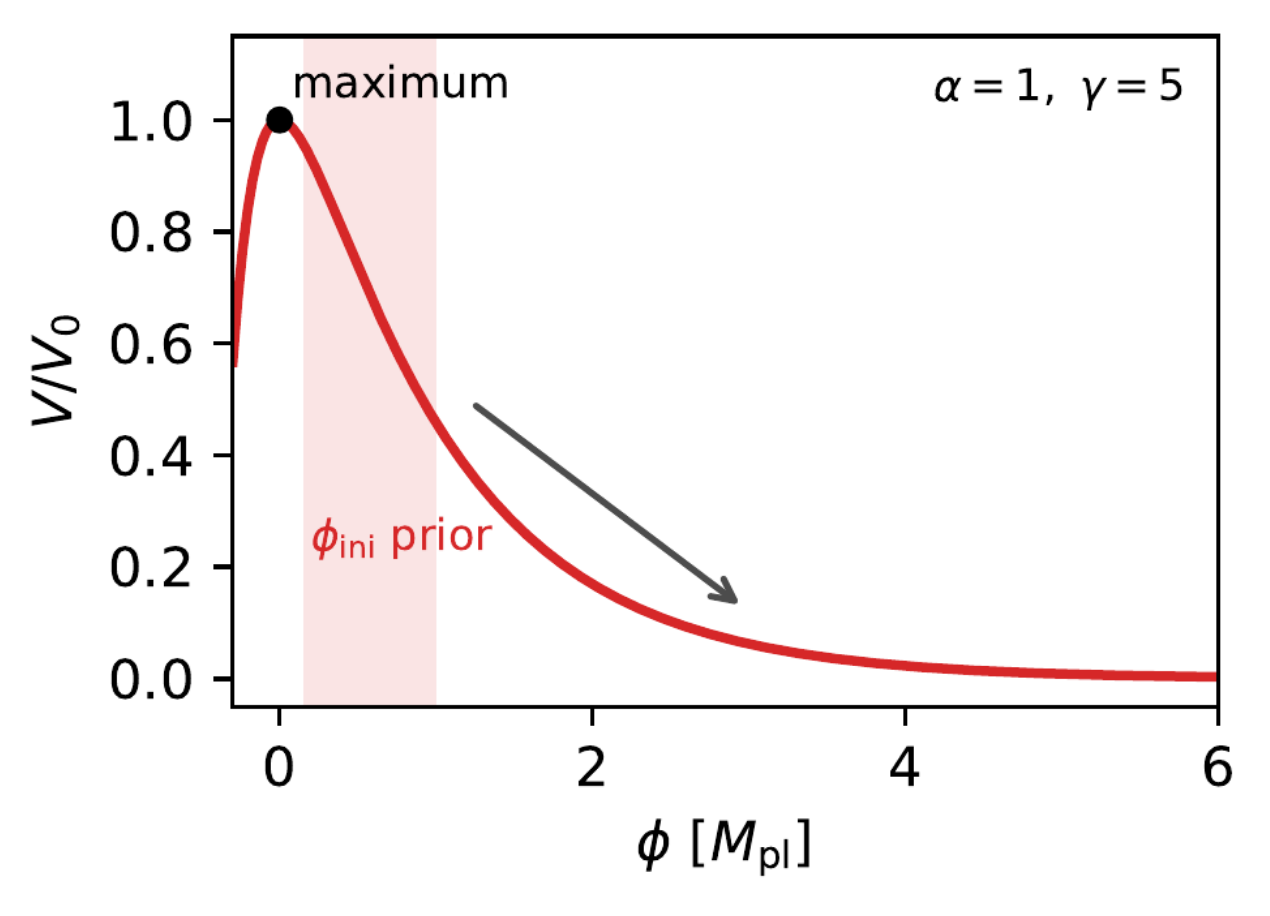}
  \caption{Shape of the double-exponential hilltop, Eq.~\eqref{eq:doubleexp},
  shown for $\alpha=1$ and $\gamma=5$, with the field starting on the slope
  beyond the maximum and rolling toward larger $\phi$. The shaded band is the
  prior range of $\phiini$.}
  \label{fig:pot_dexp}
\end{figure}

The double-exponential hilltop is,
\begin{equation}
  V(\phi) = \frac{V_0}{\gamma-\alpha}
            \left[\gamma\,e^{-\alpha\kappa\phi}
                 -\alpha\,e^{-\gamma\kappa\phi}\right],
  \qquad \gamma>\alpha>0 ,
  \label{eq:doubleexp}
\end{equation}
where $\kappa=\sqrt{8\pi G}$, with a maximum at $\phi=0$ where $V=V_0$ and
$V'=0$. Beyond the maximum the potential decreases monotonically,
approaching zero as $\phi\to\infty$: it is a runaway potential with no
minimum, and the field rolls indefinitely with larger $\phi$. This is
illustrated in Fig.~\ref{fig:pot_dexp}.This shape is motivated by the
qualitative form of the cosmon potential obtained from quantum
gravity~\cite{Wetterich2024}. Single exponentials are
the most-studied quintessence potentials because they admit exact scaling
solutions, but they can also be a limitation: a single exponential cannot both track the background early and
drive acceleration late; two slopes decouple the two regimes.

These two are dissimilar: periodic against hilltop, one scale
plus a decay constant against two logarithmic slopes, and, as
Sec.~\ref{sec:method_shooting} describes, different closure behaviour
requiring different shooting variables. Agreement between them is therefore a
non-trivial test rather than a reparametrisation.

\subsection{The dark matter-dark energy coupling}
\label{sec:model_coupling}

We take the dark matter mass  to depend on the scalar field as
$m_c(\phi)\propto e^{-\beta\kappa\phi}$, with $\beta$ a constant
dimensionless coupling~\cite{Wetterich1995,Amendola2000}. Baryons are uncoupled, so
that local fifth-force constraints are satisfied automatically. This
coupling modifies the covariant conservation equations in the dark
sector~\cite{Amendola2000},
\begin{equation}
  \nabla^\mu T^{(\phi)}_{\mu\nu} = \beta\kappa\, T^{(c)}\nabla_\nu\phi ,
  \qquad
  \nabla^\mu T^{(c)}_{\mu\nu} = -\beta\kappa\, T^{(c)}\nabla_\nu\phi ,
  \label{eq:covconserv}
\end{equation}
with $T^{(c)}_{\mu\nu}$ the dark matter energy-momentum tensor and $T^{(c)}$ its
trace.\footnote{The label $(c)$ is a superscript for tensors, to distinguish it
from the spacetime indices, and a subscript for scalar quantities such as $m_c$
and $\rho_c$.} Treating
dark matter as a pressureless perfect fluid, the background part of
Eq.~\eqref{eq:covconserv} gives
\begin{equation}
  \rho_c' + 3\mathcal{H}\rho_c = -\beta\kappa\,\rho_c\,\phi' ,
  \label{eq:rho_c_cons}
\end{equation}
where $\mathcal{H}$ is the conformal Hubble function,
which integrates to give the dark matter density,
\begin{equation}
  \rho_c(a) = \rho_{c,0}\,a^{-3}\,
              e^{-\beta\kappa\left[\phi(a)-\phi_0\right]},
  \label{eq:rho_c}
\end{equation}
and the field equation acquires a source,
\begin{equation}
  \phi'' + 2\mathcal{H}\phi' + a^2\frac{\partial V}{\partial\phi}
    = \kappa\beta a^2\rho_c ,
  \label{eq:kg}
\end{equation}
where primes denote conformal-time derivatives and a subscript $0$ the present
epoch. The model has one parameter
more than its uncoupled counterpart.

The coupling $\beta$ plays two roles~\cite{GomezValent2026}. In the
background it controls the exchange of energy between the two dark
components: as the field rolls, the dark matter mass changes, and the sign of
$\beta\phi'$ decides whether energy flows from dark matter into the field or
the reverse. Setting $\beta=0$ recovers standard cold dark matter and an
uncoupled quintessence field. In the perturbations, $\beta$ measures the
strength of a scalar fifth force between dark matter particles relative to
gravity (Sec.~\ref{sec:model_perturbations}). The energy exchange is
sensitive to the sign of $\beta$, whereas the fifth-force enhancement depends
only on $\beta^2$.

\subsection{Apparent crossing of the phantom divide}
\label{sec:model_crossing}

An observer who assumes the dark sector components are separately conserved
infers an effective dark-energy fluid~\cite{Das2006,Sola2005,Gannouji2006}
with
\begin{equation}
  \wde(a) = \frac{p_\phi}{\rho_\phi + \rho_c - \tilde\rho_c} ,
  \qquad \tilde\rho_c = \rho_{c,0}\,a^{-3} .
  \label{eq:wde}
\end{equation}
Since $w_\phi\ge-1$ for a canonical field, $\wde<-1$ requires the
denominator to fall below $\rho_\phi$, which happens when
\begin{equation}
  \beta\,\Delta\phi > 0 ,
  \qquad \Delta\phi \equiv \phi(a)-\phi_0 .
  \label{eq:crossing_condition}
\end{equation}
This is a condition on the product, not on the sign of $\beta$ alone. Whether
it selects a particular sign therefore depends on whether $\Delta\phi$ itself
changes sign with the coupling, which we determine numerically in
Sec.~\ref{sec:results_crossing}. Note that Eq.~\eqref{eq:crossing_condition}
is a background statement, independent of the perturbation sector.

\subsection{Perturbations}
\label{sec:model_perturbations}

In synchronous gauge, the default gauge of the Boltzmann code \class{} and the
one in which we implement the coupling, the coupled dark matter perturbations
obey
\begin{align}
  \delta_c' &= -\left(\theta_c + \tfrac{1}{2}h'\right)
               - \beta\,\delta\phi' ,
  \label{eq:delta_c}\\
  \theta_c' &= -\left(\mathcal{H} - \beta\phi'\right)\theta_c
               - \beta k^2\delta\phi ,
  \label{eq:theta_c}
\end{align}
which agree with the conformal-Newtonian forms of Ref.~\cite{GomezValent2026}
transformed to synchronous gauge. Two features matter for what follows. The
last term of Eq.~\eqref{eq:theta_c} acts as an attractive fifth force between dark
matter particles, which on sub-horizon scales enhances the effective
gravitational coupling to
\begin{equation}
  \frac{G_\mathrm{eff}}{G} = 1 + 2\beta^2 .
  \label{eq:geff}
\end{equation}
One factor of $\beta$ comes from the charge of the dark matter particle and
the other from the source, so the enhancement is blind to the sign of the
coupling. The drag term $\beta\phi'\theta_c$, by contrast, is linear in
$\beta$ and changes sign with it.

\section{Data and method}
\label{sec:method}

\subsection{Datasets}

We adopt the following datasets:

\textbf{BAO.} DESI DR2~\cite{DESIDR2}, using the full covariance.

\textbf{CMB.} Planck NPIPE/PR4 CamSpec TTTEEE for $\ell\ge30$~\cite{Rosenberg2022},
the 2018 low-$\ell$ TT and EE likelihoods~\cite{Planck2018V}, and the PR4
lensing likelihood~\cite{Carron2022} (marginalised).

\textbf{SNIa.} Pantheon$+$~\cite{Brout2022}, Union3~\cite{Rubin2025} and
DES\,Y5~\cite{DESY5}, each combined separately
with the above. Using all three compilations allows a direct
comparison with the DESI DR2 dark-energy results~\cite{DESIDR2}, which are
quoted for the same three samples. Our setup differs from that analysis in
two respects. First, DESI combines Planck PR4 lensing with ACT DR6 lensing,
whereas we use Planck PR4 lensing alone; this may in part add differences. Second, we neglect neutrino masses
($\sum m_\nu\simeq0$, $N_\mathrm{eff}=3.046$), whereas the DESI baseline
fixes $\sum m_\nu=0.06$\,eV, which lowers the inferred $H_0$ and $S_8$
slightly. Both choices are common to all four models compared here, so they do
not affect their relative fit, but they slightly modify the absolute results,
and our $w_0w_a$CDM baseline should be compared with Ref.~\cite{DESIDR2} with
these differences in mind.

We note that DES\,Y5 has since been recalibrated as DES-Dovekie~\cite{Popovic2025}, and that the
original calibration is known to inflate the evidence for evolving dark
energy. We retain DES\,Y5 for comparability with the published analyses we
compare against, and flag it where it matters.

\subsection{Sampled parameters and priors}

The six standard cosmological parameters
($\omega_b$, $\omega_\mathrm{cdm}$, $H_0$, $\ln10^{10}A_s$, $n_s$,
$\tau_\mathrm{reio}$) 
and the Planck nuisance parameters are sampled
throughout, in the standard $\LCDM$ way. 
We define $\omega_b\equiv\Omega_bh^2$ and $\omega_{\rm cdm}\equiv\Omega_{c}h^2$, with $h\equiv H_0/(100\,\mathrm{km\,s^{-1}\,Mpc^{-1}})$.
For the coupling we adopt a two-sided flat prior
$\beta\in\mathcal{U}[-1,1]$ in a single chain ($\mathcal{U}$ denotes a uniform distribution). This lets the data rather than
the prior decide the sign: the relative weight of the two signs is measured
within a single posterior, and $\beta=0$ lies inside the prior, so any minimum of the posterior
at $\beta=0$ is a feature of the data rather than of a prior boundary.  The price is
slower convergence, since the chains must move between the two modes
(Sec.~\ref{sec:convergence}). For the cosine potential we
sample $\log_{10}M^4\in\mathcal{U}[-9,-6]$ and
$\log_{10}f\in\mathcal{U}[-1,0]$, i.e.\ $f\in[0.1,1]\,\Mpl$; for the double-exponential,
$\alpha\in\mathcal{U}[0.01,3]$, $\gamma\in\mathcal{U}[0.1,10]$ subject to
$\gamma>\alpha$, and $\phiini\in\mathcal{U}[0.15,1.0]$. All of these priors
are flat. The upper limit on
$\phiini$ reflects the saturation of the dynamics above $\phiini\sim1$; the
lower limit keeps the field on the slope of the potential, so that it is free
to thaw. Section~\ref{sec:results_priors}
shows that the data are insensitive to $\phiini$ across this range.

\label{sec:method_shooting}
To match today's DE density we shoot on the initial field value for the cosine
potential and on the amplitude $V_0$ for the double-exponential. Shooting on
$\phiini$ fails for the double-exponential when $\alpha\lesssim0.5$. Near the
hilltop the potential is almost flat, $V'(\phiini)\approx0$, so the field
barely moves and $\Omega_\phi$ today hardly depends on $\phiini$. The amplitude
$V_0$, by contrast, sets $\Omega_\phi$ directly. The choice of shooting variable
is important: it determines which parameters carry the prior, and hence the
region of parameter space that is explored.

Theory predictions come from our own purposely modified version of \class{}~\cite{Blas2011},
sampling with \cobaya{}~\cite{Torrado2021} and
post-processing with GetDist~\cite{Lewis2025}.

\label{sec:convergence}
Each analysis uses four independent chains, which reach a Gelman-Rubin
statistic~\cite{GelmanRubin1992} of $R-1=0.010$--$0.025$. For a
bimodal posterior (Sec.~\ref{sec:results_bimodal}), where convergence is
known to be difficult to establish~\cite{Dunkley2005}, this mainly reflects differences between
chains in the relative weight of the two modes, by up to ten percentage points
in $P(\beta<0)$, rather than incomplete exploration, since every chain
recovers both modes at the same positions.

\section{Results}
\label{sec:results}

Throughout this section, every analysis combines DESI DR2 BAO and Planck PR4 CMB
data with one supernova sample, and we label each analysis only by the name of
that sample (Pantheon$+$, Union3 or DES\,Y5).

\subsection{The coupling posterior}
\label{sec:results_bimodal}

\begin{figure}[htbp]
  \centering
  \includegraphics[width=\columnwidth]{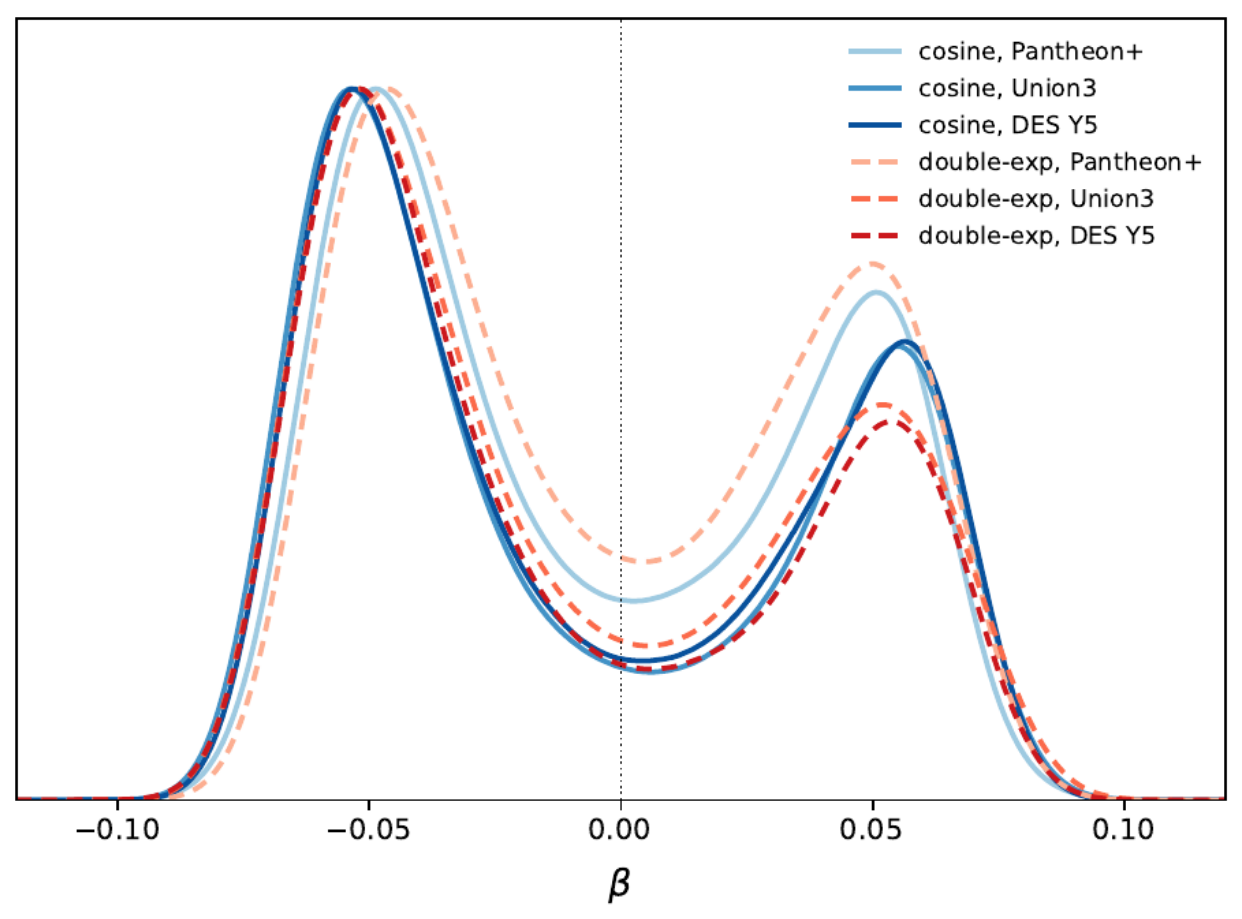}
  \caption{Marginalised posterior for the coupling, for both potentials and
  all three supernova compilations. Solid curves are the cosine potential and
  dashed the double-exponential; the dotted vertical line marks $\beta=0$.
  The posterior is bimodal in every case, with modes near $\beta\simeq\pm0.05$
  and a minimum at zero, and the six curves agree closely despite the two
  potentials having quite different geometry. Both modes persist when the
  bandwidth of the kernel density estimate is varied by a factor of nearly
  three: the trough fills in as the estimate is oversmoothed but never
  closes.}
  \label{fig:beta}
\end{figure}

The marginalised coupling posterior is bimodal for both potentials and
consistently across all three supernova compilations, with modes at
$\beta\simeq\pm0.05$ and a pronounced minimum at $\beta=0$, where the density
falls to $19$--$34\%$ of its peak value (Fig.~\ref{fig:beta}), so that
$\beta=0$ lies outside the $91$--$96\%$ credible region ($1.7$--$2.1\sigma$).
The negative
mode is the dominant one in every analysis, carrying $54$--$63\%$ of the
posterior mass (Table~\ref{tab:branch_chi2}).

The data therefore prefer a non-zero coupling magnitude. Marginalising over
both modes,
\begin{equation}
  |\beta| < 0.073 \quad (95\%\ \mathrm{CL}),
  \label{eq:beta_bound}
\end{equation}
with the posterior peaking at $|\beta|\simeq0.05$ rather than at the origin.

\begin{figure*}[htbp]
  \centering
  \includegraphics[width=\textwidth]{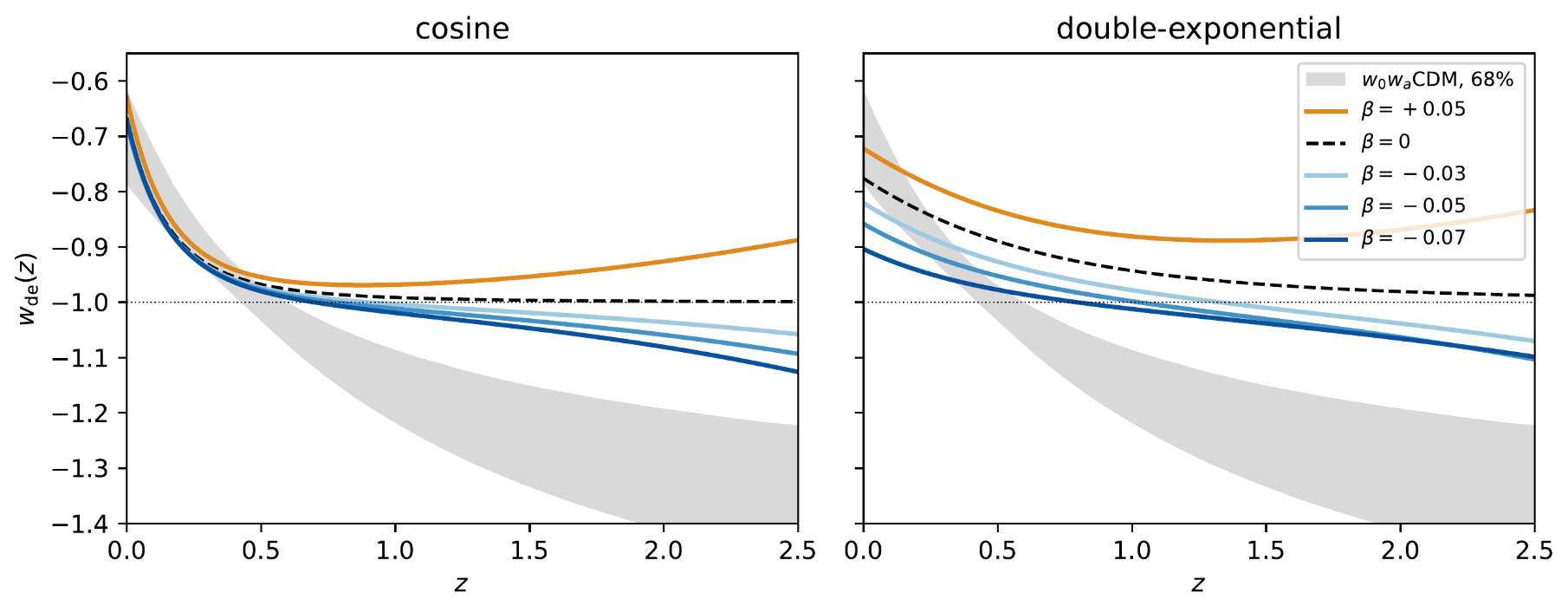}
  \caption{Effective equation of state at the Union3 best fit of each
  potential, varying only the coupling. Grey: the $68\%$ band of the
  $w_0w_a$CDM fit to the same data.}
  \label{fig:betascan}
\end{figure*}

\begin{figure*}[htbp]
  \centering
  \includegraphics[width=\textwidth]{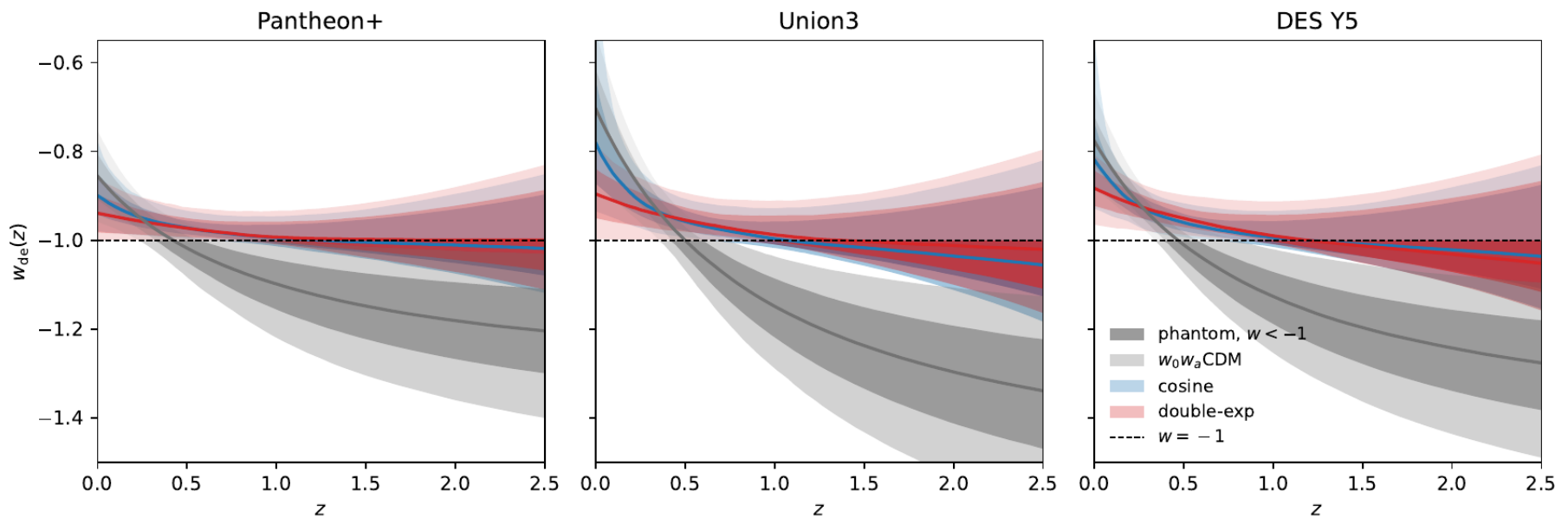}
  \caption{Effective dark-energy equation of state, reconstructed from the
  posterior of each analysis. Bands show the $68\%$ and $95\%$ credible
  regions; the portion of each band lying below $\wde=-1$ is shaded darker to
  mark the phantom region, and solid lines are the pointwise posterior
  medians. The $w_0w_a$CDM posterior fitted to identical data is shown in grey.
  Both coupled models track the phenomenological fit out to $z\simeq0.4$ and
  then depart from it. Samples on the $\beta<0$ branch cross the divide near
  $z\simeq0.9$--$1.0$, and the medians cross it near $z\simeq1.1$--$1.4$ in
  every case except the double-exponential with Pantheon$+$, whose median
  reaches $-1$ without crossing; all remain within $0.04$ of $-1$, whereas the
  CPL fit
  crosses near $z\simeq0.4$--$0.5$ and continues to fall, reaching
  $\wde\simeq-1.30$ by $z=2$ for Union3.}
  \label{fig:wz}
\end{figure*}

\label{sec:results_crossing}
The two modes are not physically equivalent. Integrating the background for
posterior samples from each branch, only the negative branch produces the
phantom crossing, at $z_\mathrm{cross}\simeq0.9$--$1.0$: $95$--$100\%$ of its
samples cross, against at most $0.2\%$ of the positive branch
(Fig.~\ref{fig:wz}). This follows from
Eq.~\eqref{eq:crossing_condition}, because the field displacement
$\Delta\phi$ is set by the potential gradient and stays negative in almost all
samples for either sign of the coupling, so that $\beta\Delta\phi>0$ requires
$\beta<0$. Figure~\ref{fig:betascan} illustrates this at the Union3 best fit
of each potential: with all other parameters fixed and the closure condition
re-solved, negative $\beta$ drives $\wde$ below $-1$ in the past, more
strongly the larger $|\beta|$, while $\beta=+0.05$ does not cross the divide.

\label{sec:results_branch_chi2}

Splitting the posterior by the sign of the coupling, the $\beta<0$ branch has
the better best fit in five of the six analyses, by
$\Delta\chi^2_\mathrm{min}=0.8$--$2.8$ (Table~\ref{tab:branch_chi2}), and is
level with the other in the sixth, the double-exponential with Pantheon$+$. The
posterior-averaged $\chi^2$, which is not affected by the numerical noise of
the minimisation, favours it in all six by $0.4$--$1.4$, with the largest
difference where $P(\beta<0)$ is largest. For the double-exponential with
DES\,Y5 the $\beta<0$ best fit lies at the lower edge of the $\phiini$ prior,
where the coupling drives the field backwards, and does not itself cross;
$95$--$97\%$ of the $\beta<0$ posterior does. The sign preference is therefore a
likelihood preference rather than a prior-volume effect, but a weak one.

\begin{table}[htbp]
  \centering
  \caption{Fit quality of the two branches, $\beta>0$ minus $\beta<0$, so
  that positive values favour the $\beta<0$ branch, which contains the
  crossing. $\chi^2_\mathrm{min}$ is from the
  noise-tolerant minimiser of Sec.~\ref{sec:results_modelcomp}, reproducible to
  $\simeq1$; $\Delta\langle\chi^2\rangle$ is the difference of posterior means.}
  \label{tab:branch_chi2}
  \begin{tabular}{llccc}
    \toprule
    potential & SN sample & $\Delta\chi^2_\mathrm{min}$
      & $\Delta\langle\chi^2\rangle$ & $P(\beta<0)$ \\
    \midrule
    cosine & Pantheon+ & $+1.75$ & $+0.43$ & $56.8\%$ \\
           & Union3    & $+1.25$ & $+0.80$ & $60.0\%$ \\
           & DES\,Y5   & $+0.84$ & $+0.63$ & $58.8\%$ \\
    \midrule
    double-exp & Pantheon+ & $-0.50$ & $+0.62$ & $54.3\%$ \\
               & Union3    & $+1.99$ & $+1.19$ & $60.9\%$ \\
               & DES\,Y5   & $+2.80$ & $+1.40$ & $63.1\%$ \\
    \bottomrule
  \end{tabular}
\end{table}

The data therefore prefer the branch that produces the phantom crossing, at
the level of roughly one unit in $\chi^2$, and this is what generates a
posterior split of $54$--$63\%$ rather than a decisive one. Between $37\%$
and $46\%$ of the allowed parameter space reproduces the DESI distance
measurements without $\wde$ ever falling below $-1$. Apparent phantom
crossing is one way for this model to accommodate the data, and on the
present evidence not a required one.

\subsection{Dependence on the potential}
\label{sec:results_potential_independence}

The two potentials give coupling constraints that are statistically
indistinguishable: the posteriors of Fig.~\ref{fig:beta} overlap almost
exactly for each supernova compilation. Quantitatively, we compare the
coupling magnitude $|\beta|$, which has a single peak, rather than $\beta$
itself, whose mean falls in the trough between the two modes. The mean of
$|\beta|$ is $0.041$--$0.047$ for both potentials and agrees to within
$0.12\sigma$ on every sample, the $95\%$ bounds on $|\beta|$ ($0.070$,
$0.073$ and $0.073$ for Pantheon$+$, Union3 and DES\,Y5) agree to within
$0.0003$, and the mode positions match to within $0.005$. This is expected if
the data respond to the coupling only through effects that are insensitive to
how the field rolls: the fifth force depends on $\beta^2$ alone, and the
crossing condition $\beta\Delta\phi>0$ depends only on the sign of
$\Delta\phi$, which is the same for both potentials. The constraint on $\beta$
is therefore a property of coupled thawing quintessence rather than of a
particular potential.

\begin{figure}[htbp]
  \centering
  \includegraphics[width=\columnwidth]{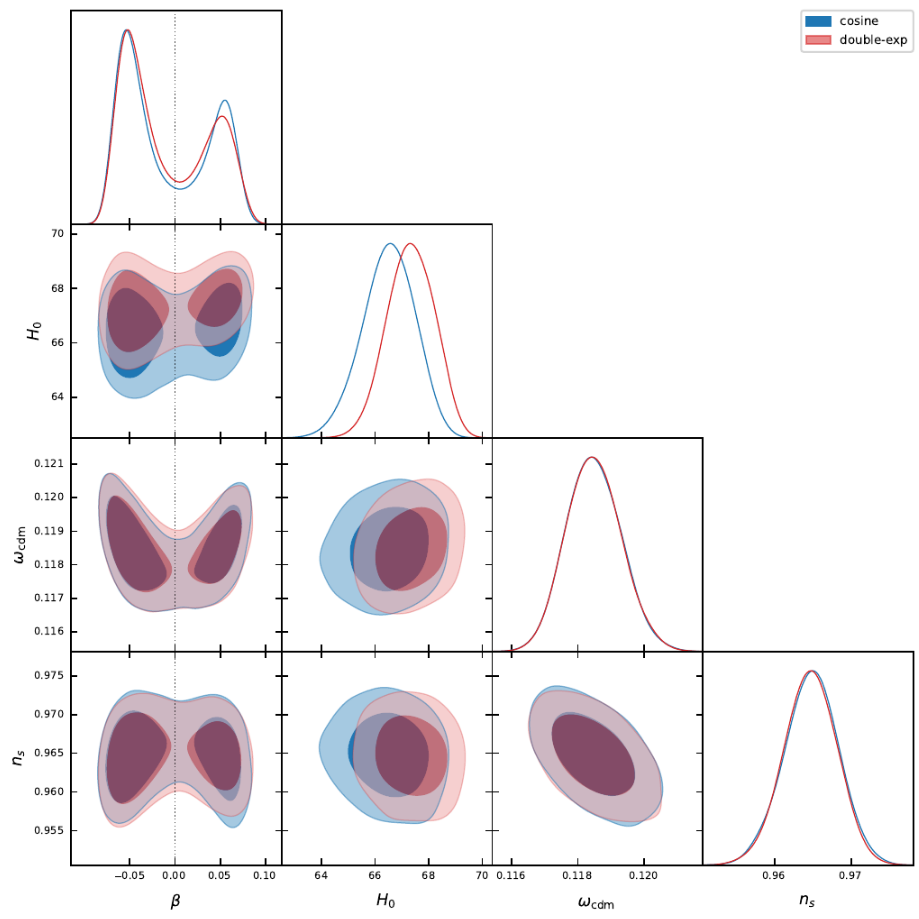}
  \caption{The two potentials compared on identical data (Union3). The
  contours overlap almost completely for $\omega_\mathrm{cdm}$ and $n_s$; the
  coupling posteriors are indistinguishable; and the only visible difference
  is a systematic offset of $\simeq0.8\ \mathrm{km\,s^{-1}Mpc^{-1}}$ in
  $H_0$, which follows from the larger departure of $w_0$ from $-1$ for the
  cosine (see text).}
  \label{fig:tri_main}
\end{figure}

The agreement extends to the growth sector. Post-processing posterior samples
through the Boltzmann code gives
$\szwelve = 0.8074$--$0.8083$ across all six analyses, a spread of $0.001$
against a per-analysis uncertainty of $0.007$, and
$S_8 = 0.825$--$0.831$. We quote $\szwelve$, the root-mean-square mass
fluctuation at a fixed scale of $12\,\mathrm{Mpc}$, rather than $\sigma_8$,
because the latter is defined at $8\,h^{-1}$Mpc and therefore refers to a
different physical scale in cosmologies with different $H_0$; this matters
here because $H_0$ differs systematically between our two potentials.

The standard cosmological parameters other than $H_0$ are unaffected by the
choice of potential, shifting by at most $0.07\sigma$ between the two; $H_0$
shifts by $0.4$--$0.9\sigma$ (Fig.~\ref{fig:tri_main}),
tracking its larger departure of $w_0$ from $-1$ (Table~\ref{tab:w0wa}): with
the CMB acoustic scale fixed, dark energy that is less negative today requires
a lower $H_0$ and higher $\Omega_m$, the same degeneracy that lowers $H_0$ in
$w_0w_a$CDM. The shift is largest for Union3, where the two values of $w_0$
differ most.

Reference~\cite{GomezValent2026} constrains the same coupling with a
Peebles-Ratra potential using an independent implementation, and finds a peak
at $|\beta|\simeq0.03$--$0.04$ lying $\gtrsim2\sigma$ from zero, an exclusion
of $\LCDM$ at no more than $2\sigma$ with Pantheon$+$ and DES-Dovekie and at
$3\sigma$ with DES\,Y5, and $\szwelve\simeq0.801$, $S_8\simeq0.812$. Our
$|\beta|$ posterior peaks somewhat higher, at $\simeq0.05$, with $\beta=0$
lying outside the $91$--$96\%$ credible region ($1.7$--$2.1\sigma$), so that the
preference for a non-zero coupling is at the same $\simeq2\sigma$ level in
both analyses; the difference in $|\beta|$ is within the posterior width. Our
exclusion levels of $\LCDM$, $1.1$--$1.4\sigma$ with Pantheon$+$ and
$2.7$--$3.2\sigma$ with DES\,Y5 (Table~\ref{tab:chi2}), are likewise of the
same size, while our $S_8$ is higher by $0.013$--$0.019$. Taken with the present
work, three thawing potentials now give a coupling magnitude of a few per
cent.

\label{sec:results_w0wa}

\begin{figure*}[htbp]
  \centering
  \includegraphics[width=\textwidth]{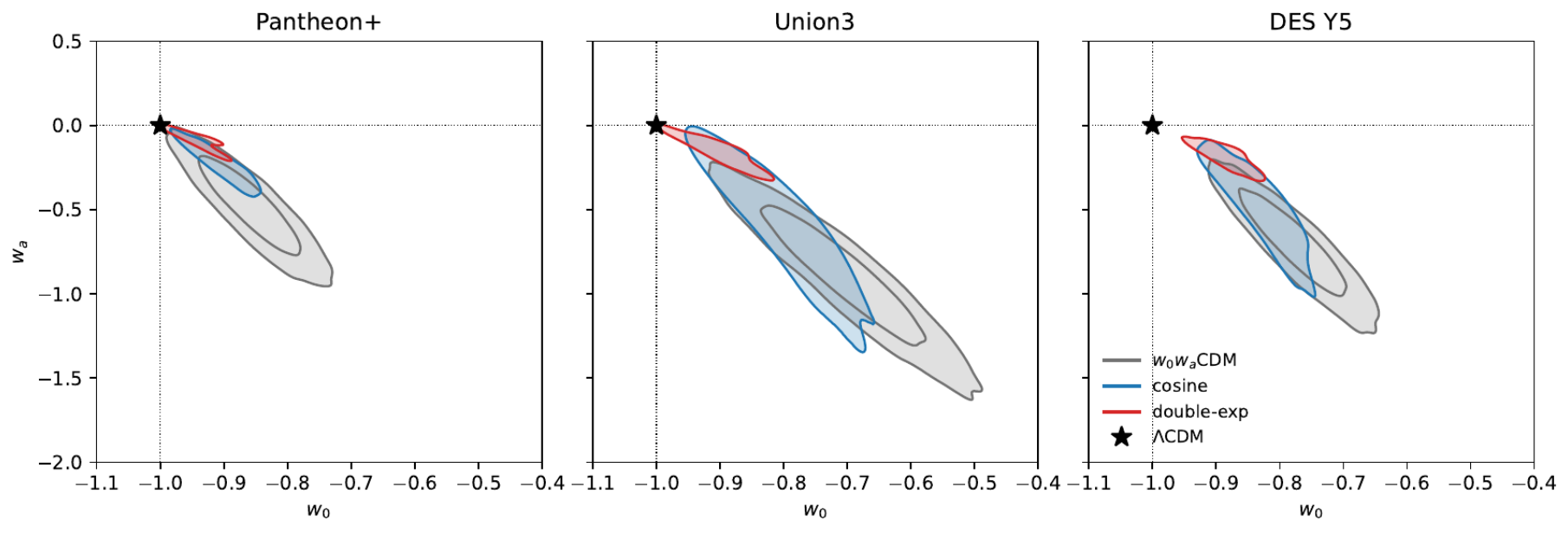}
  \caption{Effective CPL parameters of the two coupled potentials, against the
  $w_0w_a$CDM posterior fitted to identical data. Each panel combines DESI DR2
  BAO and Planck CMB data with the supernova sample indicated. Contours are $68\%$ for the
  coupled models and $68\%$ and $95\%$ for $w_0w_a$CDM; the $95\%$ region of
  the cosine extends to $w_a\simeq-3$. The star marks $\Lambda$CDM. Since the coupled models are not directly parametrized by $w_0,w_a$, these are derived from the reconstructed
  $\wde(a)$ of each posterior sample as $w_0=\wde(a{=}1)$ and
  $w_a=-\mathrm{d}\wde/\mathrm{d}a|_{a=1}$. Figure~\ref{fig:wz} shows why this
  local comparison should be read with care.}
  \label{fig:w0wa}
\end{figure*}

The one quantity on which the two potentials differ is the effective equation
of state they can reach. Table~\ref{tab:w0wa} gives
$w_0 \equiv \wde(a=1)$ and $w_a \equiv -\mathrm{d}\wde/\mathrm{d}a|_{a=1}$,
computed from the same posterior samples, together with the $w_0w_a$CDM fit to
the same data. The cosine reaches much further toward the region preferred by
the CPL fit: the CPL value of $w_a$ lies within its $68\%$ interval on Union3
and DES\,Y5 and at its edge on Pantheon$+$, whereas the double-exponential
falls short in $w_a$ by $0.4$--$0.7$ and excludes the CPL value on every
sample (Fig.~\ref{fig:w0wa}).

\begin{table}[htbp]
  \centering
  \caption{Effective CPL parameters of the two coupled models and of the
  $w_0w_a$CDM fit to the same data. Values are posterior medians with $68\%$
  intervals; the distribution of $w_a$ for the cosine is strongly skewed, with
  a long tail toward negative values.}
  \label{tab:w0wa}
  \begin{tabular}{llcc}
    \toprule
    model & SN sample & $w_0$ & $w_a$ \\
    \midrule
    cosine & Pantheon+ & $-0.90^{+0.05}_{-0.04}$ & $-0.24^{+0.12}_{-0.18}$ \\
           & Union3    & $-0.78^{+0.13}_{-0.09}$ & $-0.62^{+0.32}_{-0.77}$ \\
           & DES\,Y5   & $-0.82^{+0.08}_{-0.06}$ & $-0.48^{+0.20}_{-0.55}$ \\
    \midrule
    double-exp & Pantheon+ & $-0.94^{+0.04}_{-0.04}$ & $-0.09^{+0.06}_{-0.09}$ \\
               & Union3    & $-0.90^{+0.06}_{-0.05}$ & $-0.16^{+0.09}_{-0.12}$ \\
               & DES\,Y5   & $-0.88^{+0.04}_{-0.04}$ & $-0.18^{+0.07}_{-0.10}$ \\
    \midrule
    $w_0w_a$CDM & Pantheon+ & $-0.86^{+0.05}_{-0.05}$ & $-0.48^{+0.19}_{-0.20}$ \\
                & Union3    & $-0.70^{+0.09}_{-0.08}$ & $-0.88^{+0.26}_{-0.29}$ \\
                & DES\,Y5   & $-0.78^{+0.06}_{-0.06}$ & $-0.70^{+0.20}_{-0.21}$ \\
    \bottomrule
  \end{tabular}
\end{table}

These effective parameters describe the expansion history only near $a=1$. At
higher redshift (Fig.~\ref{fig:wz}) both coupled models cross the divide near
$z\simeq0.9$--$1.0$ and stay within $0.04$ of it, while the CPL fit crosses
near $z\simeq0.4$--$0.5$ and keeps falling, reaching $\wde\simeq-1.30$ at
$z=2$ for Union3. The cosine is therefore consistent with the present-day
value and local slope of the CPL fit but not with its high-redshift evolution,
and the double-exponential with neither: a thawing field, frozen by Hubble
friction until late times, cannot follow the continued descent the CPL
parametrisation allows.

\label{sec:results_priors}

\begin{figure*}[htbp]
  \centering
  \includegraphics[width=0.48\textwidth]{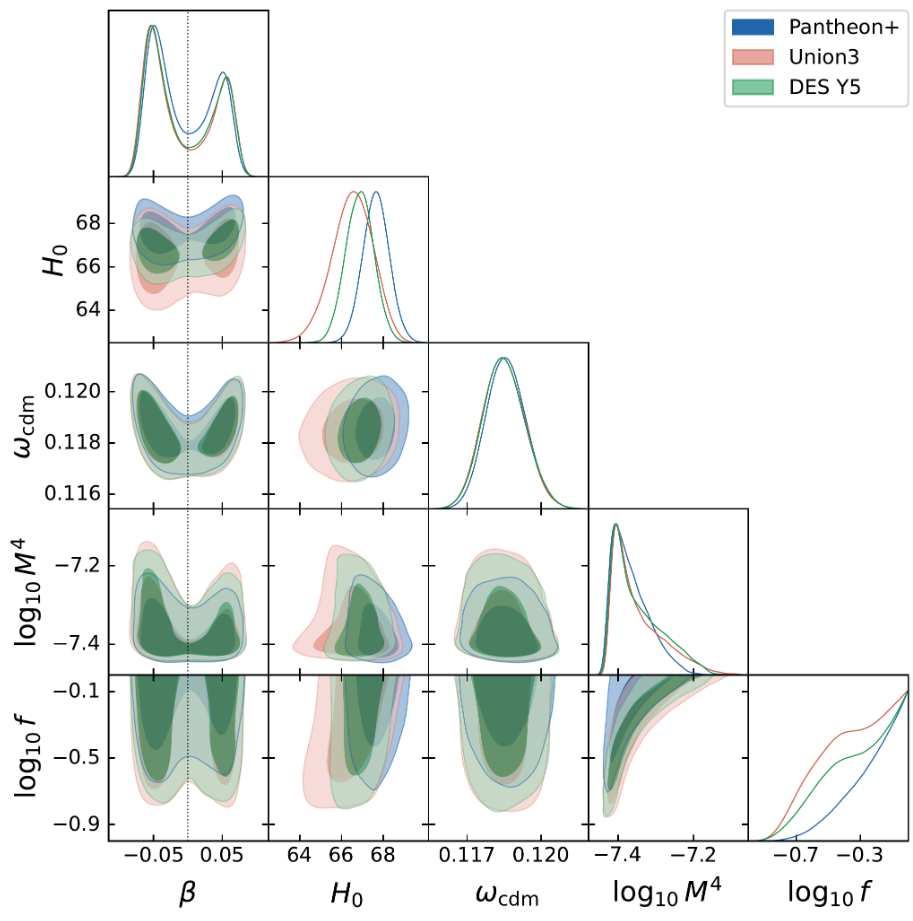}\hfill
  \includegraphics[width=0.48\textwidth]{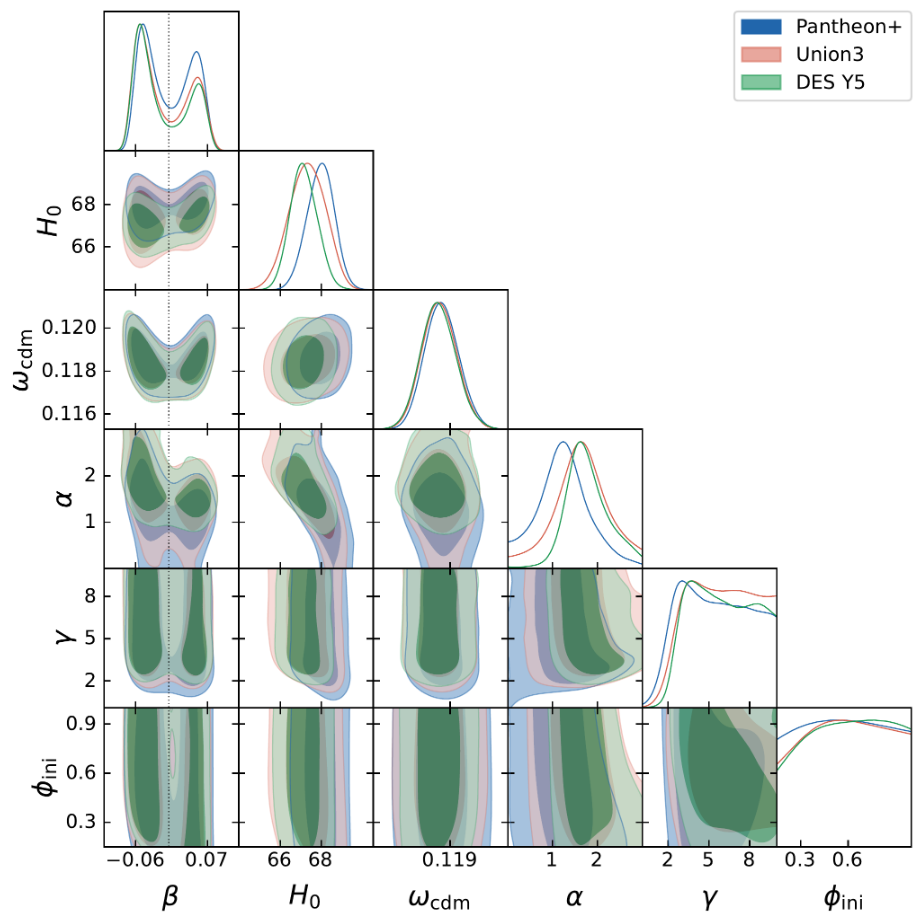}
  \caption{
  Marginalised constraints for the cosine potential (left) and the
  double-exponential (right), for the three supernova compilations. Dotted
  lines mark $\beta=0$. The bimodal coupling posterior is visible in every
  panel involving $\beta$. Of the potential parameters, only $\alpha$ is
  bounded on both sides, and only for Union3 and DES\,Y5; $\gamma$ and $\phiini$ are unconstrained or bounded on one side
  only, and $\log_{10}f$ rises toward the upper prior boundary.}
  \label{fig:triangles}
\end{figure*}

Neither potential has its shape parameters measured, as expected, since the
data constrain the late-time expansion and growth rather than the detailed form
of $V(\phi)$ (Fig.~\ref{fig:triangles}). For the double-exponential, $\phiini$
is unconstrained: its posterior density varies by at most $40\%$ across the
prior $[0.15,1.0]$, so the constraint on $\beta$ cannot be inherited from it,
which answers whether our prior choice drives the results. The steeper slope
$\gamma$ is bounded from below only, with $\gamma>1.9$, $2.4$ and $2.6$ at
$95\%$ for Pantheon$+$, Union3 and DES\,Y5, and is flat above the bound. The
shallower slope $\alpha$ is bounded from below for Union3 and DES\,Y5, with
$\alpha>0.53$ and $0.94$ at $95\%$, while for Pantheon$+$ only the upper limit
$\alpha<2.27$ survives; the lower bound appears and tightens with the strength
of the preference for evolving dark energy.

The two degeneracies have different origins. For $\gamma$ it is saturation:
once $\gamma$ exceeds a few, the second exponential in
Eq.~\eqref{eq:doubleexp} decays too fast to contribute over the field range
traversed, and $V(\phi)$ reduces to a single exponential. For $\phiini$ it is
compensation: the closure condition is enforced by shooting on $V_0$, so a
shift in the initial field value is absorbed by a shift in $V_0$. In both cases
the data respond to the effective shape of $V(\phi)$ over the range the field
traverses, not to the parameters individually.

For the cosine potential the decay constant is bounded from below only,
$f>0.20$--$0.29\,\Mpl$ ($95\%$ CL, depending on the supernova sample), with no
upper bound within the prior $f\le\Mpl$. The data therefore favour the
flat-potential regime, in which $V$ approaches a cosmological constant, a
region in some tension with the axionic Weak Gravity Conjecture requirement
$f\lesssim\Mpl$. The coupling constraints agree between the two potentials
despite these differences, which is further evidence that the constraint on
$\beta$ is not inherited from the potential.

These constraints also explain why the double-exponential falls short in $w_a$
despite its two slopes and its extra parameter: the data push $\gamma$ to
values at which the potential reduces to a single exponential. What matters is
the logarithmic slope $\lambda\equiv-V'/(\kappa V)$ felt by the field. For the
double-exponential $\lambda$ rises from zero at the maximum but can never
exceed $\alpha$, so $\wde$ departs from $-1$ only gradually; for the cosine,
$\lambda=2\tan(\phi/f)/(\kappa f)$ grows without bound as the field approaches
the minimum, allowing the rapid late-time evolution that a large $|w_a|$
requires. The extra parameter of the double-exponential thus controls a feature
the data do not constrain, while the one they respond to, a slope that steepens
at late times, is capped.

\subsection{Model comparison}
\label{sec:results_modelcomp}

\begin{table*}[htbp]
  \centering
  \caption{Minimum $\chi^2$ and, in parentheses, the improvement over
  $\LCDM$ on the same data and the corresponding exclusion level of $\LCDM$.
  The latter follows from the likelihood-ratio test, assuming Wilks' theorem
  with $2$, $3$ and $4$ additional parameters for $w_0w_a$CDM, the cosine and
  the double-exponential respectively, converted to a number of standard
  deviations as for a univariate Gaussian, following Ref.~\cite{GomezValent2026}.
  The lower block gives the difference in the deviance information criterion,
  with the effective number of parameters estimated as half the posterior
  variance of $\chi^2$ (see text). All values are obtained with the same procedure:
  BOBYQA in its noisy-objective mode with global restarts, seeded at the best
  chain sample. The likelihood carries numerical noise of $\simeq1$ in
  $\chi^2$ from the shooting step, which sets the precision of each entry.}
  \label{tab:chi2}
  \begin{tabular}{lccc}
    \toprule
    model & Pantheon+ & Union3 & DES\,Y5 \\
    \midrule
    $\LCDM$      & $12389.31$ & $11012.76$ & $12633.24$ \\
    $w_0w_a$CDM  & $12382.14\ (-7.17;\ 2.2\sigma)$ & $10999.23\ (-13.53;\ 3.3\sigma)$
                 & $12616.65\ (-16.59;\ 3.7\sigma)$ \\
    cosine       & $12384.17\ (-5.14;\ 1.4\sigma)$ & $11002.84\ (-9.92;\ 2.3\sigma)$
                 & $12617.97\ (-15.27;\ 3.2\sigma)$ \\
    double-exp   & $12384.22\ (-5.09;\ 1.1\sigma)$ & $11004.08\ (-8.68;\ 1.8\sigma)$
                 & $12618.89\ (-14.36;\ 2.7\sigma)$ \\
    \midrule
    \multicolumn{4}{l}{$\Delta\mathrm{DIC}$ relative to $\LCDM$} \\
    $w_0w_a$CDM  & $-2.9$ & $-8.5$ & $-14.0$ \\
    cosine       & $-2.4$ & $-4.9$ & $-10.6$ \\
    double-exp   & $-1.7$ & $-2.7$ & $-8.2$ \\
    \bottomrule
  \end{tabular}
\end{table*}

Both coupled models improve on $\LCDM$ for every supernova sample, recovering
between $64\%$ and $92\%$ of the improvement achieved by $w_0w_a$CDM, which
fits best in every case. Between the two potentials, the cosine fits better on
Union3 and DES\,Y5, and the two are level on Pantheon$+$, where their effective
$(w_0,w_a)$ are also closest (Table~\ref{tab:w0wa}). Since the
double-exponential has one more free parameter, the cosine is preferred by
the Akaike information criterion on all three samples, by
$\Delta\mathrm{AIC}\simeq2$--$3$. This preference is consistent but weak,
and information criteria do not decisively separate any of the four models.
The exclusion levels of $\LCDM$ are $1.4$--$3.2\sigma$ for the cosine and
$1.1$--$2.7\sigma$ for the double-exponential, against $2.2$--$3.7\sigma$ for
$w_0w_a$CDM, rising from Pantheon$+$ to DES\,Y5 with the strength of the
preference for evolving dark energy. They are approximate: the data constrain only
some of the potential parameters, so we count all of them as additional, which
gives the conservative (lower) values.

We also compute the deviance information criterion~\cite{Spiegelhalter2002},
$\mathrm{DIC}=\langle\chi^2\rangle+p$, used in the DESI
analyses~\cite{Lodha2025}. Its standard effective number of parameters,
$p_D=\langle\chi^2\rangle-\chi^2(\bar\theta)$, requires $\chi^2$ at the
posterior mean $\bar\theta$, which for our bimodal posterior lies between the
modes where the fit is poor; we therefore use $p_V=\mathrm{var}(\chi^2)/2$
instead~\cite{Gelman2013}, and have checked that the two definitions agree to
within $1.2$ for $\LCDM$ and $w_0w_a$CDM. The resulting $\Delta\mathrm{DIC}$
(Table~\ref{tab:chi2}) is $-1.7$ to $-10.6$ for the coupled models, against
$-2.9$ to $-14.0$ for $w_0w_a$CDM, and favours the cosine over the
double-exponential on every sample. The effective number of additional
parameters is $1.2$--$2.4$ for the coupled models, fewer than their nominal
three and four, which confirms that the exclusion levels above are
conservative.

\subsection{The Hubble constant and the growth of structure}
\label{sec:results_H0}

The coupling correlates positively with $H_0$ (correlation coefficient $+0.2$
to $+0.3$), but the bound on $|\beta|$ limits the effect to
$\lesssim0.5\ \mathrm{km\,s^{-1}Mpc^{-1}}$. We obtain
$H_0=66.5$--$68.0\ \mathrm{km\,s^{-1}Mpc^{-1}}$ across the six analyses,
below the $\LCDM$ value of $\simeq68.3$ for the same data, so that, like
$w_0w_a$CDM, the coupled models mildly worsen rather than relieve the Hubble
tension.

\label{sec:results_growth}

\begin{figure*}[htbp]
  \centering
  \includegraphics[width=\textwidth]{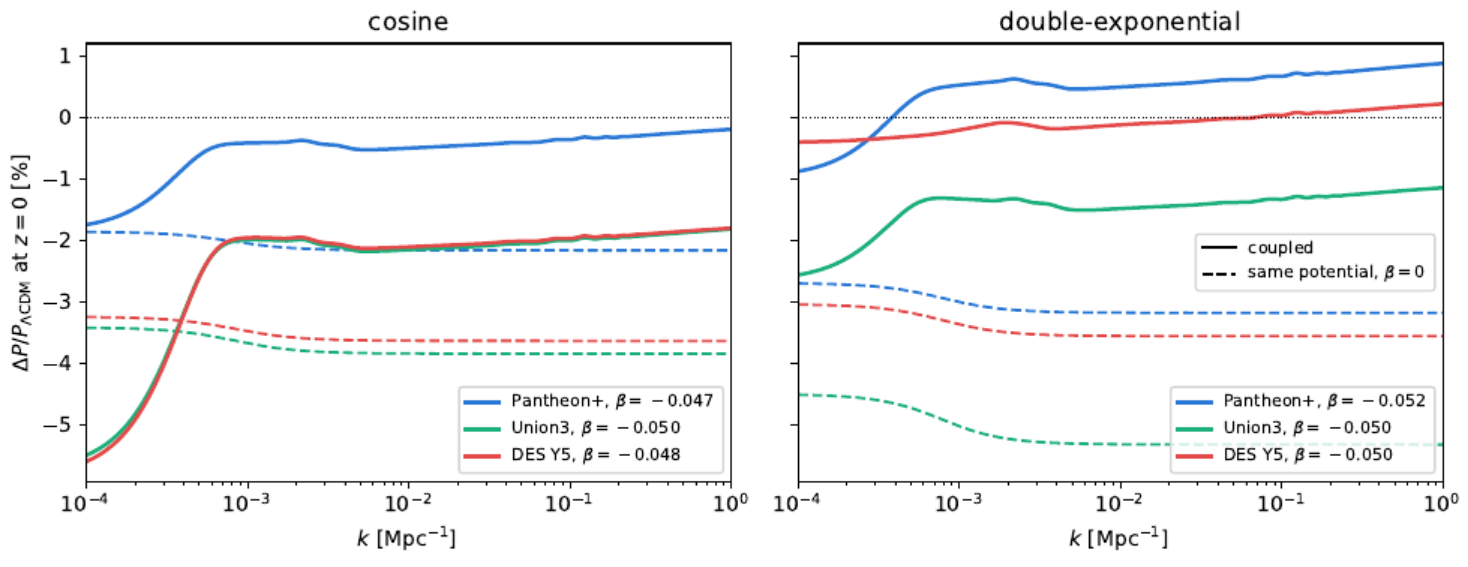}
  \caption{Linear matter power spectrum at $z=0$ for the best fits
  with $\beta<0$, relative to $\LCDM$ with the same cosmological parameters. Solid
  curves are the coupled models and dashed curves the same potential with
  $\beta=0$. The thawing field alone suppresses growth by $2$--$5\%$, and the
  fifth force restores most of it, so that the two effects largely cancel. Below $k\simeq10^{-3}\,\mathrm{Mpc}^{-1}$
  the fifth force is cut off beyond the mediator's Compton wavelength.}
  \label{fig:pk}
\end{figure*}

Varying only the coupling at fixed cosmology, $\sigma_8$ responds as
\begin{equation}
  \frac{\Delta\sigma_8}{\sigma_8}
    = -0.013\,\beta \;+\; 2.62\times 2\beta^2 ,
  \label{eq:sigma8_fit}
\end{equation}
so at $|\beta|=0.05$ the sign-dependent linear term is twenty times smaller
than the quadratic one, and growth data constrain $|\beta|$ but carry almost
no information about its sign. Figure~\ref{fig:pk} shows the matter power
spectrum at the best fits. The thawing field alone suppresses $P(k)$ by
$2$--$5\%$ on scales $k\gtrsim10^{-3}\,\mathrm{Mpc}^{-1}$, and the fifth force
restores most of this, leaving the coupled models within about $2\%$ of
$\LCDM$ at the same cosmological parameters; the compensation is more
complete for the double-exponential. The suppression by the thawing field and
the enhancement by the fifth force therefore largely cancel, so the net effect
on clustering is small. On larger scales the fifth force is cut off beyond the
mediator's Compton wavelength.

The coupled models give $S_8=0.825$--$0.831$, above the $\LCDM$ value of
$0.818$--$0.820$ for the same data. The clustering amplitude itself is not
enhanced, $\sigma_8=0.807$--$0.818$ against $0.817$, because the fifth force
largely compensates the suppression of growth by the thawing field
(Fig.~\ref{fig:pk}); the higher $S_8$ instead follows from the larger
$\Omega_m$ that accompanies the lower $H_0$, as for $w_0w_a$CDM. The coupled
models therefore do not relieve the $S_8$ tension. The observational prospects
for the growth sector are discussed in Sec.~\ref{sec:future_lensing}.

\section{Outlook}
\label{sec:future}

The coupling enters most observables quadratically: the fifth force depends on
$\beta^2$, and the sign-dependent contribution to distances is also of second
order, so that at $|\beta|=0.05$ the quadratic part of the growth response
exceeds the linear part by a factor of twenty (Eq.~\eqref{eq:sigma8_fit}).
Sensitivity to $|\beta|$ therefore scales as the square root of the
measurement precision: halving the uncertainty requires a fourfold improvement
in the data.

\label{sec:future_lensing}
For the magnitude, the main prospect is Stage-IV weak lensing. Inverting
Eq.~\eqref{eq:sigma8_fit}, a $2\sigma$ detection of the fifth force requires
$|\beta|\gtrsim\sqrt{\sigma(S_8)/(2.62\,S_8)}$. Current lensing measurements,
with $\sigma(S_8)\simeq0.02$, reach only $|\beta|\gtrsim0.09$, whereas
\emph{Euclid} and LSST\,Y10, with $\sigma(S_8)\simeq0.004$ and $0.003$, reach
$|\beta|=0.043$ and $0.037$, below the mode positions we find. These
thresholds assume a fixed background, so the partial compensation by the
thawing field (Fig.~\ref{fig:pk}) will weaken them, and they rely on
\textsc{halofit}; a quantitative analysis requires the non-linear treatment of
Refs.~\cite{Tudes2025}. Later DESI releases and the LSST supernova
sample will also tighten the background constraint, which currently
dominates, subject to the same square-root scaling.

None of these determines the sign of the coupling, and by
Eq.~\eqref{eq:geff} no growth measurement can. An observable linear in
$\beta$ exists in this model: because baryons are uncoupled, the drag term
$-\beta\kappa\phi'\theta_c$ of Eq.~\eqref{eq:theta_c} gives dark matter and
baryons a relative velocity that is linear in the coupling and reverses with
it. Measurements of the cosmological Euler equation with multi-tracer galaxy
surveys~\cite{Castello2025,ZhengEuler2025} probe exactly this quantity, and
since the apparent phantom crossing is controlled by the sign of
$\beta\Delta\phi$, they are the most direct route to settling it.

Finally, all our analyses neglect neutrino masses. The bound on $\sum m_\nu$
depends on how the dark-energy equation of state evolves~\cite{Feng2026} and,
in interacting dark energy models, on the form and sign of the
coupling~\cite{Li2026nu}. Freeing the neutrino mass could therefore broaden or
shift the $\beta$ posterior and change the relative weight of the two
branches, offering a second, indirect handle on the sign.

\section{Comparison with previous constraints}
\label{sec:literature}

Reported constraints on this coupling span a wide range of significances and
disagree on the sign. Much of this spread is explained by differences in how
the two signs are treated in the analysis rather than by differences in the
data.

Ref.~\cite{GomezValent2026} analyses $\beta>0$ and $\beta<0$ as two distinct
cases, each with its own chain, obtaining
$\beta = +0.035^{+0.016}_{-0.009}$ and $\beta = -0.030^{+0.012}_{-0.018}$
with Planck\,PR4 $+$ DESI\,DR2 $+$ Pantheon$+$, at
$\chi^2_\mathrm{min} = 12389.92$ and $12390.01$ respectively, a difference of
$0.09$. Sampling a two-sided prior in a single chain, we recover the same
structure as one bimodal posterior. The two descriptions are consistent;
indeed their near-equality of $\chi^2_\mathrm{min}$ between the two signs is
precisely the statement that the likelihood is nearly invariant under
$\beta\to-\beta$, which Eq.~\eqref{eq:sigma8_fit} makes quantitative. The
bimodal form makes explicit what the comparison of separate runs implies.

Ref.~\cite{deSouza2026} models both dark components as axion-like fields,
with a cosine potential for dark energy and a field-dependent coupling arising
from their mixing, and restricts the coupling to $\beta\ge0$, since in their
construction the crossing requires that sign. Their improvement over $\LCDM$,
$\Delta\chi^2\simeq-9$ to $-14$, and their late crossing of the phantom
divide, near $z\simeq1.5$, are similar to ours, while their Bayesian evidence
is inconclusive. A one-sided posterior, however, cannot test the sign, and the
resulting exclusion of the uncoupled model is not directly comparable with a
two-sided analysis.

Ref.~\cite{Li2026} reports evidence for a non-vanishing coupling at the
$3$--$5\sigma$ level, using additional small-scale CMB data from ACT and
SPT-3G and the DES-Dovekie supernovae. For coupled quintessence with an
inverse power-law potential and the same coupling as ours they find
$\beta\simeq0.05$, close to our mode positions, but with a one-sided prior
$\beta\ge0$. For a field rolling toward larger $\phi$ this selects the
non-crossing branch, consistent with their finding that the model does not
require a phantom crossing, and it cannot reveal the second mode; their
single-peaked posterior is therefore not in tension with our bimodal one.
Their additional small-scale CMB data may sharpen the magnitude of the
coupling, although by Eq.~\eqref{eq:geff} growth observables respond to
$\beta^2$ and cannot by themselves distinguish the signs. We have not
attempted to reproduce their analysis.

\section{Conclusions}
\label{sec:conclusions}

We set out to test whether coupled thawing quintessence can reproduce the DESI
DR2 result for any thawing potential, or whether this depends on the potential
chosen. Our conclusions are as follows.

\emph{Coupled quintessence reproduces the qualitative DESI result.} For both
potentials and every supernova sample, the effective dark-energy equation of
state today lies above $-1$, with medians $w_0=-0.78$ to $-0.94$, and on the
weakly favoured negative branch it crosses into the phantom regime in the past, as in
the DESI reconstructions, without a phantom field. This evolution improves the
fit over $\LCDM$ by $\Delta\chi^2=5$--$15$, an exclusion of $\LCDM$ at
$1.1$--$3.2\sigma$, recovering $64$--$92\%$ of the improvement achieved by
$w_0w_a$CDM. What the coupled models do not reproduce
is the depth of the phantom phase: $\wde$ crosses $-1$ later, near
$z\simeq0.9$--$1.0$ rather than $z\simeq0.4$--$0.5$, and stays within $0.04$
of it, whereas the CPL fit reaches $\wde\simeq-1.30$ by $z=2$. The departure
from $-1$ that DESI prefers is thus captured at low redshift, but not its
steep descent at high redshift.

\emph{The coupling constraint is potential-independent.} Two geometrically
distinct thawing potentials give constraints agreeing to within
$0.15\sigma$ on every dataset, $|\beta| < 0.073$ at $95\%$, and growth
predictions agreeing to $\szwelve = 0.807\pm0.008$. Taken with the
Peebles-Ratra results of Ref.~\cite{GomezValent2026}, three thawing
potentials now give the same answer.

\emph{The accessible equation of state is not.} The cosine potential is
consistent with the present-day value and local slope of the CPL fit to the
same data, with the CPL $w_a$ inside its $68\%$ interval (on DES\,Y5,
$w_a=-0.48^{+0.20}_{-0.55}$ against $-0.71$), while the double-exponential falls short in $w_a$ by $0.4$--$0.7$ on
every sample, because the slope its field feels can never exceed the shallower
exponent $\alpha$, whereas the slope of the cosine steepens without bound as
the field approaches the minimum. This is reflected weakly in the fit: with one parameter fewer,
the cosine is preferred on all three samples by
$\Delta\mathrm{AIC}\simeq2$--$3$, most clearly where the evidence for
evolving dark energy is strongest.
Potential-independence of the coupling constraint does not imply
potential-independence of the expansion history.

\emph{Neither the Hubble nor the $S_8$ tension is relieved.} The coupled
models infer $H_0=66.5$--$68.0\ \mathrm{km\,s^{-1}Mpc^{-1}}$, below the $\LCDM$
value for the same data, and $S_8=0.825$--$0.831$, above it; the higher $S_8$
follows from the larger $\Omega_m$ that accompanies the lower $H_0$ rather than
from enhanced clustering, since the fifth force largely compensates the
suppression of growth by the thawing field.

\emph{The sign of the coupling is not determined.} The posterior is bimodal
with modes at $\beta\simeq\pm0.05$; only the negative branch produces the
phantom crossing, and it is favoured by only $\Delta\chi^2\simeq1$. Between
$37\%$ and $46\%$ of the allowed parameter space reproduces the DESI
distances without $\wde$ ever falling below $-1$. Apparent phantom crossing
is therefore one way for this model to accommodate the data, not a
requirement of it.

\emph{Determining the sign requires observables linear in $\beta$.} Growth
observables respond to $\beta^2$ and are blind to the sign by construction.
Because baryons are uncoupled while dark matter is not, the two components
acquire a relative velocity that is linear in the coupling, and measurements
of the cosmological Euler equation through multi-tracer galaxy surveys
\cite{Castello2025,ZhengEuler2025} probe a quantity to which the present
constraint is structurally blind.

\emph{The coupling is a physically motivated explanation worth testing.} It
produces an apparent phantom crossing with a canonical scalar field, avoiding
the instabilities of a genuinely phantom one, and the data place its magnitude
at $|\beta|\simeq0.05$ with $\beta=0$ disfavoured at $\simeq2\sigma$. Our
results agree with the DESI evidence for evolving dark energy: in every
analysis the effective equation of state today lies above $-1$, and $\LCDM$ is
disfavoured at $1.1$--$3.2\sigma$, somewhat below the $2.8$--$4.2\sigma$
reported for $w_0w_a$CDM~\cite{DESIDR2}. There has been some criticism of
this evidence, both for its dependence on the priors and
parametrisation~\cite{Patel2024} and for its sensitivity to the supernova
sample and calibration~\cite{CortesLiddle2025,Popovic2025}, and the preference we find is
likewise weakest with Pantheon$+$ and strongest with DES\,Y5. Upcoming data can
separate these possibilities. If the signal is an artefact of priors or
calibration, it should weaken as the data improve and the supernova samples
are recalibrated; if the coupling is physical, Stage-IV weak lensing could
detect the associated fifth force at $|\beta|\gtrsim0.04$, and multi-tracer
measurements of the Euler equation can test its sign directly.

Coupled thawing quintessence can therefore reproduce the DESI preference
qualitatively, recovering most of the improvement of the CPL fit for both
potentials, but current data fix neither the depth of the phantom phase nor
the sign of the coupling.

\begin{acknowledgments}
We thank C.~Wetterich for suggesting the double-exponential potential.
Computations were performed on the ITP Heidelberg cluster.
LA acknowledges support by DFG under Germany’s Excellence Strategy EXC 2181/1 - 390900948 (the Heidelberg STRUCTURES Excellence Cluster), under Project 554679582 ``GeoGrav: Cosmological Geometry and Gravity with nonlinear physics'', and  by Tamkeen under the NYUAD Research Institute grant CG014.
\end{acknowledgments}

\appendix

\section{Parameter constraints}
\label{app:params}

Table~\ref{tab:params} lists the marginalised constraints on all sampled parameters,
and on the derived quantities used in the text, for the six analyses.

\begin{table*}[htbp]
  \centering
  \caption{Marginalised constraints on the sampled and selected derived
  parameters: posterior mean and standard deviation, or $95\%$ limits for
  parameters bounded on one side only. Because the coupling posterior is
  bimodal, we quote the magnitude $|\beta|$, with its mean and $95\%$ upper
  limit. Derived parameters are computed from 1000 posterior samples per
  analysis. Each column combines DESI DR2 BAO and Planck CMB data with the
  supernova sample indicated.}
  \label{tab:params}
  \begin{tabular}{lcccccc}
    \toprule
    & \multicolumn{3}{c}{cosine} & \multicolumn{3}{c}{double-exponential} \\
    \cmidrule(lr){2-4}\cmidrule(lr){5-7}
    parameter & Pantheon$+$ & Union3 & DES\,Y5 & Pantheon$+$ & Union3 & DES\,Y5 \\
    \midrule
    $\omega_b$ & $0.02221\pm0.00013$ & $0.02222\pm0.00013$ & $0.02221\pm0.00013$
               & $0.02221\pm0.00013$ & $0.02222\pm0.00013$ & $0.02222\pm0.00013$ \\
    $\omega_\mathrm{cdm}$ & $0.1186\pm0.0008$ & $0.1185\pm0.0008$ & $0.1185\pm0.0008$
                          & $0.1186\pm0.0008$ & $0.1185\pm0.0008$ & $0.1184\pm0.0008$ \\
    $H_0$ & $67.65\pm0.64$ & $66.52\pm0.99$ & $66.87\pm0.67$
          & $67.97\pm0.67$ & $67.32\pm0.88$ & $67.15\pm0.66$ \\
    $\ln10^{10}A_s$ & $3.039\pm0.014$ & $3.039\pm0.014$ & $3.039\pm0.014$
                    & $3.038\pm0.014$ & $3.039\pm0.014$ & $3.039\pm0.014$ \\
    $n_s$ & $0.9645\pm0.0035$ & $0.9648\pm0.0036$ & $0.9647\pm0.0035$
          & $0.9644\pm0.0035$ & $0.9647\pm0.0035$ & $0.9648\pm0.0037$ \\
    $\tau_\mathrm{reio}$ & $0.0546\pm0.0070$ & $0.0548\pm0.0071$ & $0.0550\pm0.0072$
                         & $0.0544\pm0.0071$ & $0.0549\pm0.0070$ & $0.0551\pm0.0071$ \\
    $|\beta|$ & $0.042\pm0.019$ & $0.047\pm0.019$ & $0.046\pm0.019$
              & $0.041\pm0.019$ & $0.044\pm0.020$ & $0.045\pm0.019$ \\
              & $<0.070$ & $<0.073$ & $<0.073$ & $<0.070$ & $<0.073$ & $<0.073$ \\
    $\log_{10}M^4$ & $-7.37\pm0.05$ & $-7.35\pm0.07$ & $-7.35\pm0.07$ & -- & -- & -- \\
    $\log_{10}f$ & $>-0.55$ & $>-0.71$ & $>-0.68$ & -- & -- & -- \\
    $\alpha$ & -- & -- & -- & $<2.27$ & $1.64\pm0.59$ & $1.78\pm0.47$ \\
    $\gamma$ & -- & -- & -- & $>1.90$ & $>2.40$ & $>2.63$ \\
    $\phiini$ & -- & -- & -- & unconstrained & unconstrained & unconstrained \\
    \midrule
    $\Omega_m$ & $0.308\pm0.006$ & $0.318\pm0.009$ & $0.314\pm0.006$
               & $0.305\pm0.006$ & $0.310\pm0.008$ & $0.312\pm0.006$ \\
    $\sigma_8$ & $0.815\pm0.010$ & $0.807\pm0.012$ & $0.810\pm0.010$
               & $0.818\pm0.010$ & $0.814\pm0.012$ & $0.812\pm0.011$ \\
    $S_8$ & $0.826\pm0.009$ & $0.831\pm0.010$ & $0.829\pm0.009$
          & $0.825\pm0.009$ & $0.828\pm0.009$ & $0.828\pm0.010$ \\
    $\szwelve$ & $0.807\pm0.008$ & $0.808\pm0.008$ & $0.808\pm0.008$
               & $0.808\pm0.008$ & $0.808\pm0.008$ & $0.808\pm0.008$ \\
    \bottomrule
  \end{tabular}
\end{table*}

\bibliographystyle{apsrev4-2}
\bibliography{refs}

\end{document}